\documentclass[reprint, superscriptaddress, aps, prl]{revtex4-2}

\usepackage{amsmath,amssymb}
\usepackage{graphicx}
\usepackage{bm}
\usepackage{times}
\usepackage{xcolor}
\usepackage{lineno}
\usepackage{textcomp, gensymb}
\usepackage{hyperref}
\usepackage{xspace}
\hypersetup{colorlinks=true, citecolor=blue, urlcolor=gray, linkcolor=cyan}

\newcommand{\CaWO}{CaWO$_{4}$ }

\newcommand{\VASP}{\textsc{VASP}\xspace}
\newcommand{\MILADY}{\textsc{MILADY}\xspace}
\newcommand{\LAMMPS}{\textsc{LAMMPS}\xspace}
\newcommand{\CRAB}{\textsc{CRAB}\xspace}
\newcommand{\CEvNS}{\textsc{CE}$\nu$\textsc{NS}\xspace}

\definecolor{asparagus}{rgb}{0.53, 0.66, 0.42}

\begin{document}

\title{Calculation of crystal defects induced in \CaWO by 100 eV displacement cascades using a linear Machine Learning interatomic potential}

\newcommand{\IRFU}{\affiliation{IRFU, CEA, Universit\'e Paris-Saclay, 91191 Gif-sur-Yvette, France}}
\newcommand{\SRMP}{\affiliation{Service de recherche en Corrosion et Comportement des Matériaux, SRMP, Universit\'e Paris-Saclay, CEA, 91191 Gif Sur Yvette, France }}

\author{G.\,Soum-Sidikov}\IRFU\SRMP
\author{J.\,P.\,Crocombette}\email{Corresponding author: Jean-Paul.CROCOMBETTE@cea.fr}\SRMP
\author{M.\,C.\,Marinica}\SRMP
\author{C. Doutre}\IRFU
\author{D. Lhuillier}\IRFU
\author{L. Thulliez} \IRFU


\date{\today}
\begin{abstract}
We determine the energy stored in the crystal defects induced by $\mathcal{O}(10-100)$\,eV nuclear recoils in low-threshold \CaWO cryogenic detectors. A Machine Learning interatomic potential is developed to perform molecular dynamics simulations. We show that the energy spectra expected from Dark Matter and neutrino coherent scattering are affected by the crystal defects and we provide reference predictions. We discuss the special case of the spectrum of nuclear recoils induced by neutron capture, which could offer a unique sensitivity to the calculated stored energies.
\end{abstract}

\maketitle


%
Cryogenic low-threshold detectors are at the crossroads of two vast experimental programs: the search for light Dark Matter (DM) and the study of Coherent Elastic Neutrino-Nucleus Scattering (\CEvNS). Despite abundant evidence of a predominantly “dark” component to the mass of the universe \cite{BERTONE2005279}, no direct interaction with ordinary matter has yet been observed. The strongest constraints have been placed on the elastic scattering of DM with mass $\ge$1 GeV \cite{Billard:2021uyg}. Lighter masses are now being considered, implying lower nuclear recoils in the detectors at the $\mathcal{O}(10\--100)$\,eV scale. Access to low detection thresholds also paves the way for the study of \CEvNS, experimentally demonstrated for the first time in 2017 \cite{akimov2017observation}. The precise study of this new neutrino-matter interaction offers opportunities for original tests of the Standard Model at low energies~\cite{Billard_2018}. 

Detection thresholds at the 10 eV scale have recently been demonstrated with Ge and Si cryogenic detectors with masses of the order of several grams \cite{PhysRevD.99.082003,EDELWEISS:2022ktt,SuperCDMS:2018mne,SuperCDMS:2020aus}. Here, we focus on cryogenic detectors made of \CaWO crystals equipped with a Transition Edge Sensor to detect the phonon (heat) signal induced by the nuclear recoils. This technology was initially developed by the CRESST experiment \cite{Strauss:2017woq,CRESST:2019jnq} for DM searches and is now applied to the detection of \CEvNS by the NUCLEUS experiment \cite{Strauss:2017cuu,NUCLEUS:2019igx}. As the elastic scattering of low speed DM particles and low energy neutrinos are coherent processes, their cross-sections are respectively proportional to the square of the number of nucleons and neutrons in the target nuclei, motivating the use of crystals containing heavy atoms like tungsten. The trade-off is a very low energy nuclear recoil of hundreds of eV or lower. Such a low energy range means that we need to study in detail the atomic-scale material effects that can affect the phonon signal from cryogenic detectors.
In particular, part of the energy deposited in the target material can be stored in crystal defects created along the displacement cascade, affecting the energy scale of the detector. Molecular dynamics~(MD) simulations of this phenomenon have already been performed with empirical potentials for some commonly used detector materials \cite{PhysRevLett.120.111301,PhysRevD.106.063012} not including \CaWO\!. In this paper, we present MD calculations based on a new Machine Learning~(ML) interatomic potential developed for this study. 
Bypassing complexity walls, ML empowers simulations to achieve atomic-level fidelity at the ab initio level, dramatically reducing computational costs.
The predicted DM and \CEvNS recoil energy spectra are then expressed as a function of the detected energy, taking into account the energy stored in crystal defects. 
We discuss as well the calibrated nuclear recoils induced by neutron capture on tungsten \cite{Thulliez:2020esw,CRAB:2022rcm,CRAB:2023kuz} as a potential experimental validation of these calculations.

%
To design the  \CaWO ML potential we first built a database of atomic configurations using Density Functional Theory (DFT) calculations. Computations were performed using the \VASP code \cite{RN7468} with the Generalized Gradient Approximation in the Perdew Burke Erzenhof form (GGA-PBE) \cite{RN839} functional. This fast functional allowed us to compute a large dataset, which would have been out of reach with more precise but much heavier functionals (e.g. hybrid ones). However we have checked our main GGA-PBE results against those obtained with the Heyd-Scuseria-Ernzerhof (HSE06) \cite{RN6205} hybrid functional. Calculations were done in $2\times2\times2$ and $3\times3\times3$ supercells of \CaWO containing 96 and 324 atoms respectively, at the $\Gamma$ point. We especially considered the energies of close interstitial-vacancy Frenkel Pairs (FPs), that are expected to be the main defects created by low energy displacement cascades. We systematically searched for the positions of interstitials using a regularly spaced grid of positions every 0.7~\AA{}, with a 1.8~\AA{} lower threshold on inter-atomic positions. FP structures were relaxed down to a threshold of $2\times10^{-3}$~eV/\AA{}. We eventually found a few inequivalent structures for tungsten, calcium and oxygen FPs, with energies listed in table \ref{tab:formation_energies}.
\begin{table}[]
    \caption{Formation energies (eV) of different type of Frenkel Pairs in \CaWO (first column) computed with the PBE functional for two different sizes of supercell (columns 2 and 3). The fourth column shows results after a further relaxation with the HSE06 functional. These energies include a correction for elastic effects due to the finite supercells sizes \cite{RN7985}. "Unstable" denotes a configuration which happens to form a stable defect in a [$2\times2\times2$] box but relaxes back to perfect positions in larger boxes.}
    \label{tab:formation_energies}
\begin{ruledtabular}
    \begin{tabular}{lccc}
         & PBE [$2\times2\times2$] & PBE [$3\times3\times3$] & HSE06 $\left[2\times2\times2\right]$ \\
       \hline
       W          &   5.56    &   7.05  &  5.89   \\
                  &   6.83    &   7.30  &  7.28   \\
                  &   8.07    &   10.59  &  8.60   \\
                  &   8.58    &   11.07  &  9.08   \\
\hline
      Ca       &  5.02    &  5.60  & 5.38 \\
               &  5.11    &  5.19  &  5.50   \\
\hline
       O       &  3.16    &    unstable   & unstable \\
               &  4.74    &  4.84  &  5.05   \\
               &  4.96     &  4.87   &  5.24   \\
               &  5.07     &  5.14   &  5.41   \\
               &  5.22     &  5.28   &  5.41   \\
    \end{tabular}
\end{ruledtabular}

\end{table}
PBE energies are quite close to the HSE06 values, giving confidence in the results later obtained with the ML potential. The database contains all the relaxed FP structures as well as some structures extracted from their minimization runs. It is complemented by excerpts from 2000~K MD runs for a perfect crystalline box and a box with one tungsten FP. We also considered boxes with a short-distance pair of atoms, including in the database both their initial configuration and the structures relaxed under the constraint of fixed close pair interatomic distance. All these additional structures were obtained in $2\times2\times2$ boxes.

Then we designed the ML interatomic potential by employing a hybrid descriptor within the framework of a linear ML (LML) model, akin to the methodology outlined in \cite{RN8699}. The local atomic environment is encapsulated into a direct sum of an accurate radial 2-body descriptor \textbf{z\textsubscript{a }}and a many-body descriptor \textbf{x}\textsubscript{a}. The former is quick to evaluate, while the latter can be numerically costly.
Both descriptors depict the neighborhood $n(a)$ of a\textsuperscript{th} atom within a cutoff distance of 4.8 \AA{}. \textbf{x}\textsubscript{a} utilizes bispectrum SO(4) descriptors \cite{RN9034, RN9035, RN8699, RN9032, RN9033}, expanding the local density $\rho_a$ of the neighborhood $n(a)$ of a\textsuperscript{th} atom,
\[ \rho_a(r) = \sum_{i \in n(a)} w_{t(i)} \delta\left(r-r_i\right)\]
\noindent
in hyperspherical 4D harmonics, where $i \in n(a)$ runs over atoms in the neighborhood of $a$. We limit the kinetic moment of 4D harmonics to j\textsubscript{max} = 4. 
The chemical species are differentiated through three channels $c$, which is equivalent to have three local atomic densities $\rho_a^c(r)$ \textit{i.e.} to have three collections of  weights $w_{t(i)}^c$, where $t(i)$ represents the type of atom $j$. The first channel assigns the three weights $w_{t(i)}^{c = 1}$ equal to 1.0, whilst for the last two channels are 0.7, 0.45 and 1.52 for Ca, O and W respectively, and 0.8, 0.3 and 1.98. 

This approach is reminiscent of that employed in \cite{RN9037} and corresponds to the case K = 3 in modern tensorial decomposition \cite{RN9038}. This descriptor inherently includes a 4-body descriptor of the atomic neighborhood, whereas the radial distance projection is not as refined \cite{RN9039}. This gap in information is filled by the descriptor \textbf{z\textsubscript{a}}, following the 2-body interaction kernel described in \cite{RN9035, RN8699}.

The advantages of using hybrid descriptors are twofold: (i)~the lack of information in the many-body descriptor is compensated for by the information in the first, and (ii) for very short distances, below an 1.3~\AA{} inner cutoff distance, the many-body interaction is switched off and the 2-body interaction is naturally coupled to the Ziegler Biersack Littmark (ZBL) potential \cite{RN795} for distances shorter than 1~\AA{}. Between 1.3~\AA{} and 1~\AA{}, a buffer region is established to ensure continuity up to the second derivatives between the ZBL and the kernel 2-body potential. These features are crucial for the accurate simulation of high-energy recoils. Indeed during recoils, interatomic distances can become so short that interatomic repulsions are not described by non full-electron DFT codes such as \VASP. The complete database comprises a total of 348 configurations. Energies, force vectors on each atom, and stress components for all the boxes were fitted using LML linear regression of the descriptors employing the \MILADY package \cite{RN9039, RN8699}. We ultimately achieved a Mean Average Error of 0.474~eV and 0.131~eV/\AA{} for energies and atomic forces respectively.

We modeled recoil events using the \LAMMPS code \cite{lammps} as the \MILADY plugin. The nearly cubic simulation box is $8.535\times8.535\times9.208\,\text{nm}^3$ and contains 49152 atoms. In the context of application to cryodetectors, extremely low temperatures (of the order 10 mK) have to be considered. Still, atoms are not at rest due to their zero point energy. To account for these residual vibrations in the Newtonian mechanics framework of MD, we considered the classical atomic system to experience a (classical) temperature such that the atomic motions are of the same order as those of the corresponding quantum system. Various correspondences exist between classical and quantum temperatures on atomic systems. We chose the simple general rule stated in~\cite{RN1316} that the classical temperature equivalent of a quantum zero temperature is $\frac{3}{8}$ of the Debye temperature of the crystal under consideration. This renormalization of temperature explains even the experimentally observed low-temperature migration of defects in complex energy landscapes \cite{Arakawa2020}.
The Debye temperature of \CaWO being 354~K \cite{RN8896}, that amounts to 133~K. The simulation box was thus initially thermalized at this temperature. 
Note that advanced treatment of quantum vibration behavior of solids exist (e.g. Path Integral Molecular Dynamics \cite{RN6689}) but are designed to explore the thermodynamical equilibrium properties of matter at very low temperature and are not applicable to vastly out of equilibrium processes such as sudden atomic recoils.

After thermalization, a tungsten atom initially at the center of the box is accelerated with a specific kinetic energy and direction, then subsequent atomic movements are followed for 8~ps, with a variable time-step and a 0.01~\AA{} threshold on the maximum atomic displacement between two consecutive steps. Initial recoil energies of 20, 40, 60, 81, 112 and 160~eV were considered. The last three energies correspond to the main calibration peaks expected from the application of the \CRAB calibration method to \CaWO \cite{Thulliez:2020esw} and cover the region of interest for the study of the DM and \CEvNS. The box is large enough so that the temperature increase at the end of the simulation (about 8~K for 112~eV) is negligible. The box is eventually quenched to reach the local energy minimum after the cascade, which, by difference with the energy of the initial defect-free box gives access to the stored energy after the recoil event.

\begin{figure}[tb]
   \includegraphics[width=\linewidth, trim={2cm 1.8cm 0cm 1.8cm},clip]{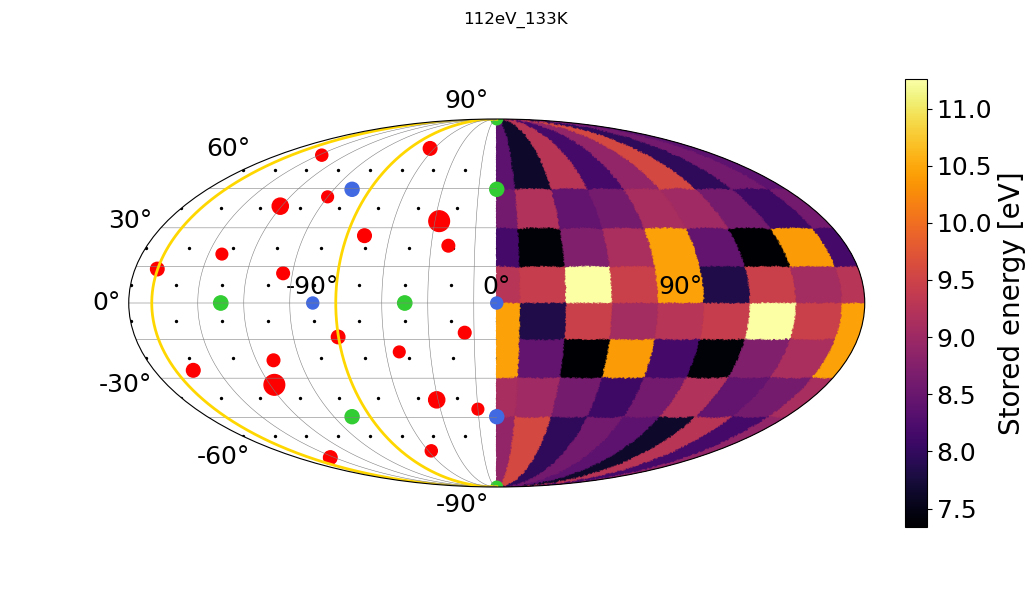}
    \caption{Mollweide projection of the ball centered at the Primary Knock-on Atom (PKA) initial location. Left part: crystal structure in a radius of 6~\AA. The color of the disks indicate the type of atom: Calcium (green), Tungsten (blue) and Oxygen (red) whereas the size of the disk is inversely proportional to the distance to the PKA. The black dots show the sampled directions of recoil. The area enclosed by the yellow line illustrates the elementary region to be covered based on the \(\overline{4}\) point symmetry. Right: energy stored in crystal defects for all initial directions of a 112~eV nuclear recoil.}
    \label{fig:Mollweide}
\end{figure}

\begin{figure}[ht]
   \includegraphics[width=\linewidth]{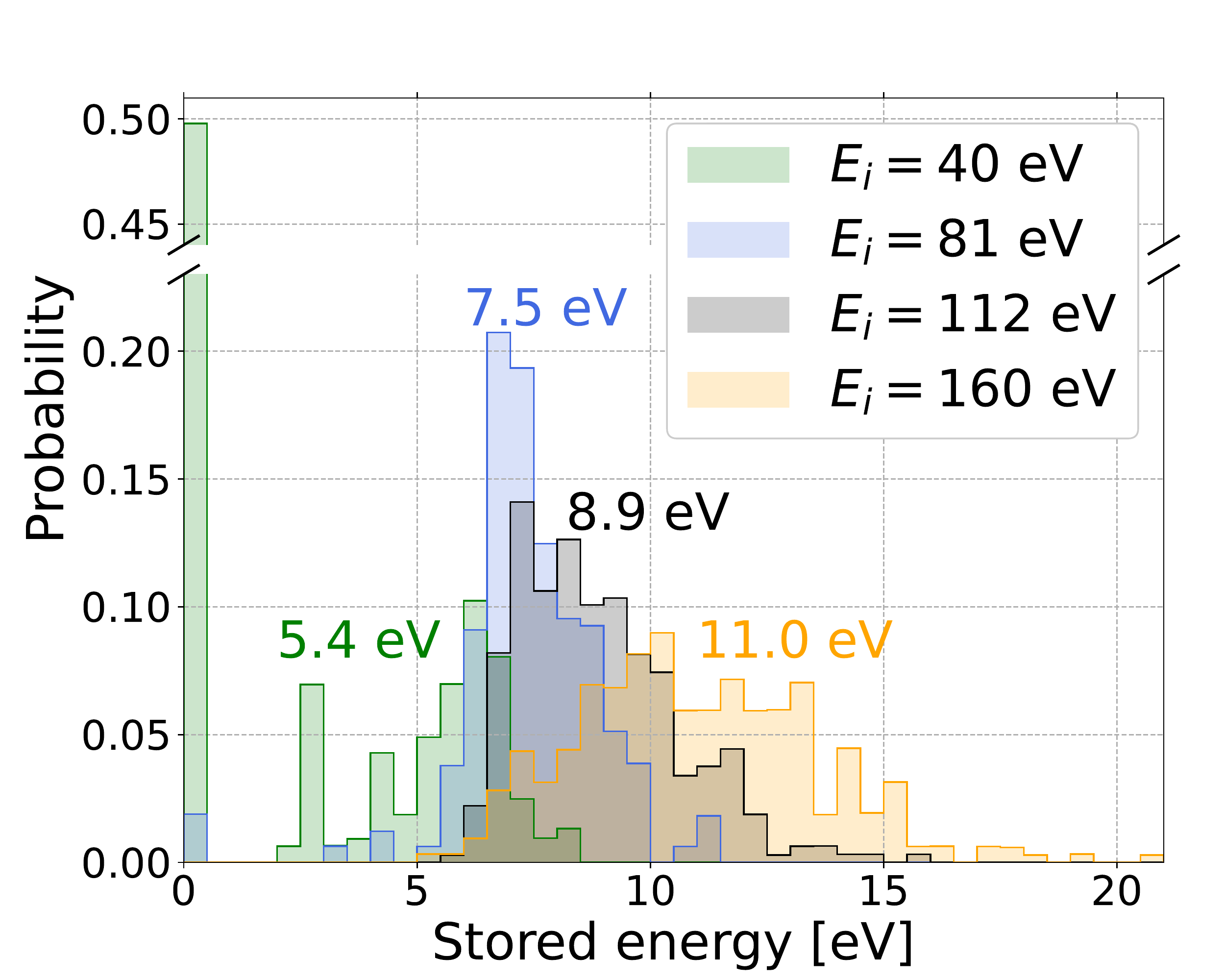}
    \caption{Predicted distribution of energy stored in \CaWO crystal defects for different recoil energies of interest: 40~eV (green, 320 configurations), 81~eV recoils (blue, 160 configurations), 112~eV recoils (gray, 320 configurations) and 160~eV recoils (orange, 320 configurations). The number next to each distribution indicates the average value of non-zero stored energies. 
    }
    \label{fig:StoredEnergies}
\end{figure}

The tungsten site has a \(\overline{4}\) point symmetry that allows sampling only one fourth of the solid angle to map it entirely. We did so, considering 32 directions sampling nearly regularly the solid angle, as illustrated in the left panel of figure \ref{fig:Mollweide}. For each direction 10 initial starting points extracted regularly from the last 10~ps of the thermalization run were considered as initial configurations (5 initial positions were actually considered for the 80~eV recoils and only 1 for 20 and 60~eV). The average over the initial starting configurations of the stored energies for a 112 eV recoil is plotted in the right panel of figure \ref{fig:Mollweide}. No clear relation appears between the surrounding of the W atom in the recoil direction and the energy stored in such events. Gathering all starting configurations and directions we eventually obtained the distributions of stored energies shown in figure \ref{fig:StoredEnergies}, with each sampled direction weighted by the covered solid-angle fraction. Approximately 10\% of the initial recoil energy is stored in crystal defects. However, the existence of threshold displacement energies implies a departure from an exact scaling with initial energy. For instance the average energy stored for nuclear recoils of 160~eV (11.0~eV) is lower than twice the energy stored for 81~eV recoils (7.5~eV). Also the smooth shape of the distributions rejects the hypothesis of sharp transitions between few defect configurations. This is confirmed by inspection of the atomic positions after the simulated displacement cascades, which shows amorphization of the medium around the recoil trajectory. 

\begin{figure}[tb]
   \includegraphics[width=\linewidth]{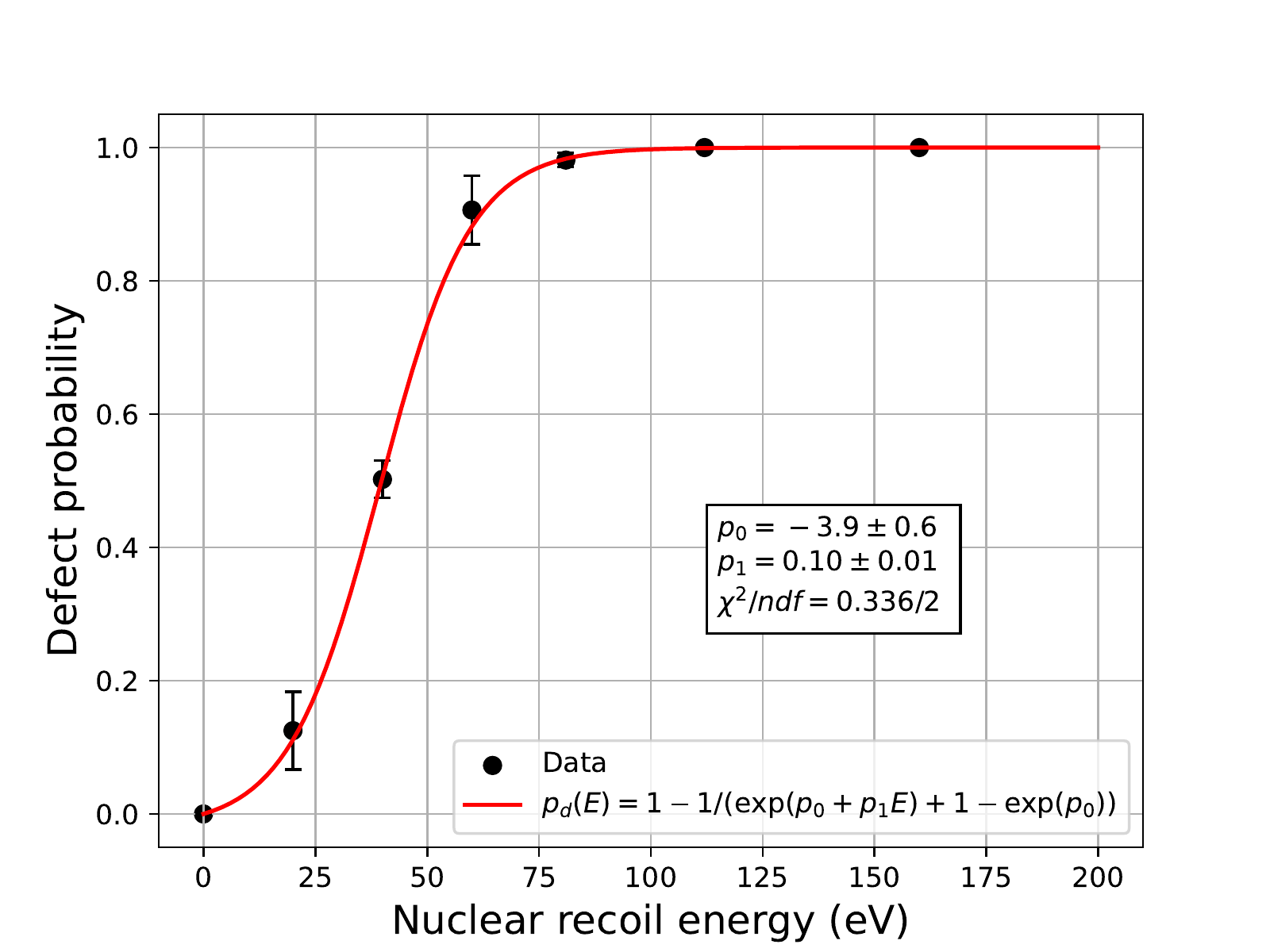}
    \caption{Evolution of the defect probability versus the initial recoil energy (black points) and its fit by a Fermi function (red curve).}
    \label{fig:p0}
\end{figure}

\begin{figure}[tb]
   \includegraphics[width=\linewidth]{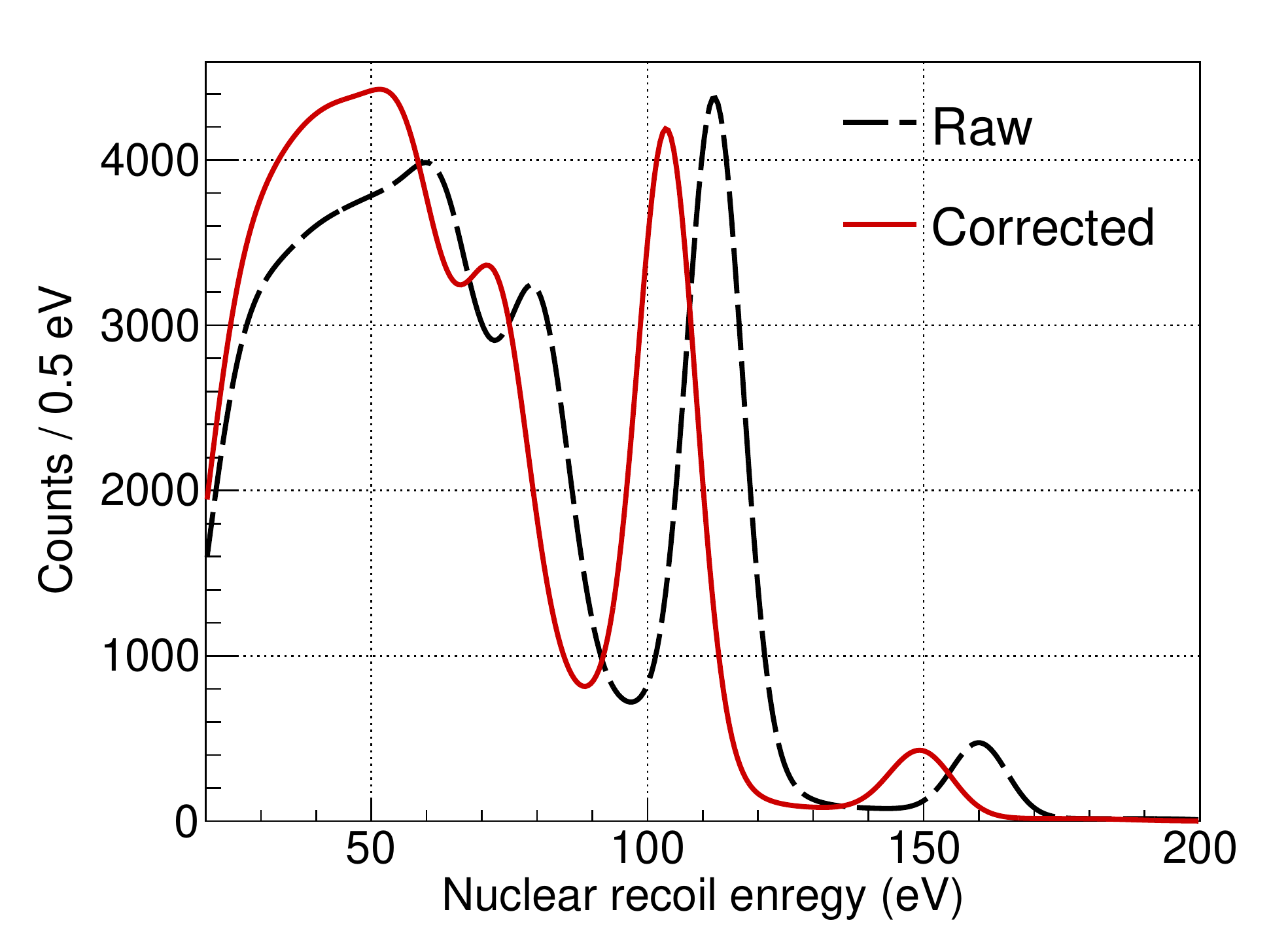}
    \caption{Recoil spectra induced by neutron captures in a 0.75~g \CaWO cryogenic detector with a 5~eV energy resolution: as initially predicted in \cite{CRAB:2023kuz} (black dashed line) and corrected for the energy stored in lattice defects (red solid line).}
    \label{fig:crab_spectrum}
\end{figure}

%
To our knowledge, such detailed calculations of displacement cascades in \CaWO are unprecedented. In order to estimate their impact on various continuous nuclear recoil spectra without having to increase the number of heavy MD simulations, we interpolate between the computed distributions of stored energy. 
Our approach consists in first fitting our results for each initial recoil energy with a shape model and then fit the dependence of the model parameters on the initial recoil energy to obtain a continuous evolution of the stored energy distributions. The uncertainty of this procedure was estimated by testing several shape models (regular or asymmetric Gaussian functions) as well as several functions to describe the evolution of the shape model parameters (first or second order polynomials and Gaussian).
We found no significant impact of the choice of a given model on our convolution calculations. 
Finally, for a given initial nuclear recoil, the shape model is normalized to the defect probability $p_d$, inferred from the relative amount of simulated configurations that shows a non-zero stored energy. We observe a smooth evolution of $p_d$ from 0 to 1 in the considered range of initial recoil energies, well fitted by a Fermi function (see figure \ref{fig:p0}). 
Any prediction of recoil spectrum should be convoluted by this continuous set of distributions to correct for the energy stored in the crystal defects. We discuss here the impact of this correction on three examples of predicted recoil spectra corresponding to ongoing or planned measurements with \CaWO cryogenic detectors. The first example is the recoil spectrum expected from exposure of a \CaWO crystal to a flux of thermal neutrons. This process has been proposed as a new and accurate calibration method based on the mono-energetic recoils of tungsten atoms induced by radiative neutron capture \cite{Thulliez:2020esw}. The three main calibration peaks expected at 81, 112 and 160~eV recoil energy are clearly visible in figure \ref{fig:crab_spectrum} as well as their shift toward $\approx$10\% lower visible energies after correcting for crystal defects. 
The measurement proposed in \cite{Thulliez:2020esw,CRAB:2023kuz} by coupling a dilution refrigerator and a beam of thermal neutrons is expected to happen next year at the TRIGA reactor in Vienna. The statistical accuracy of the position of the peaks should reach the sub-eV level after few days of data taking, providing a unique insight into the atomic-scale processes affecting the detector signal at very low energy. For instance the above-mentioned non-linearity of the detector response between 81 and 160 eV resulting from the formation of crystal defects could be tested. Furthermore this study could be performed as a function of the direction of the tungsten recoil, tagged by the detection in coincidence of the high energy $\gamma$-ray responsible for the nuclear recoil.

\begin{figure}[tb]
   \includegraphics[width=\linewidth]{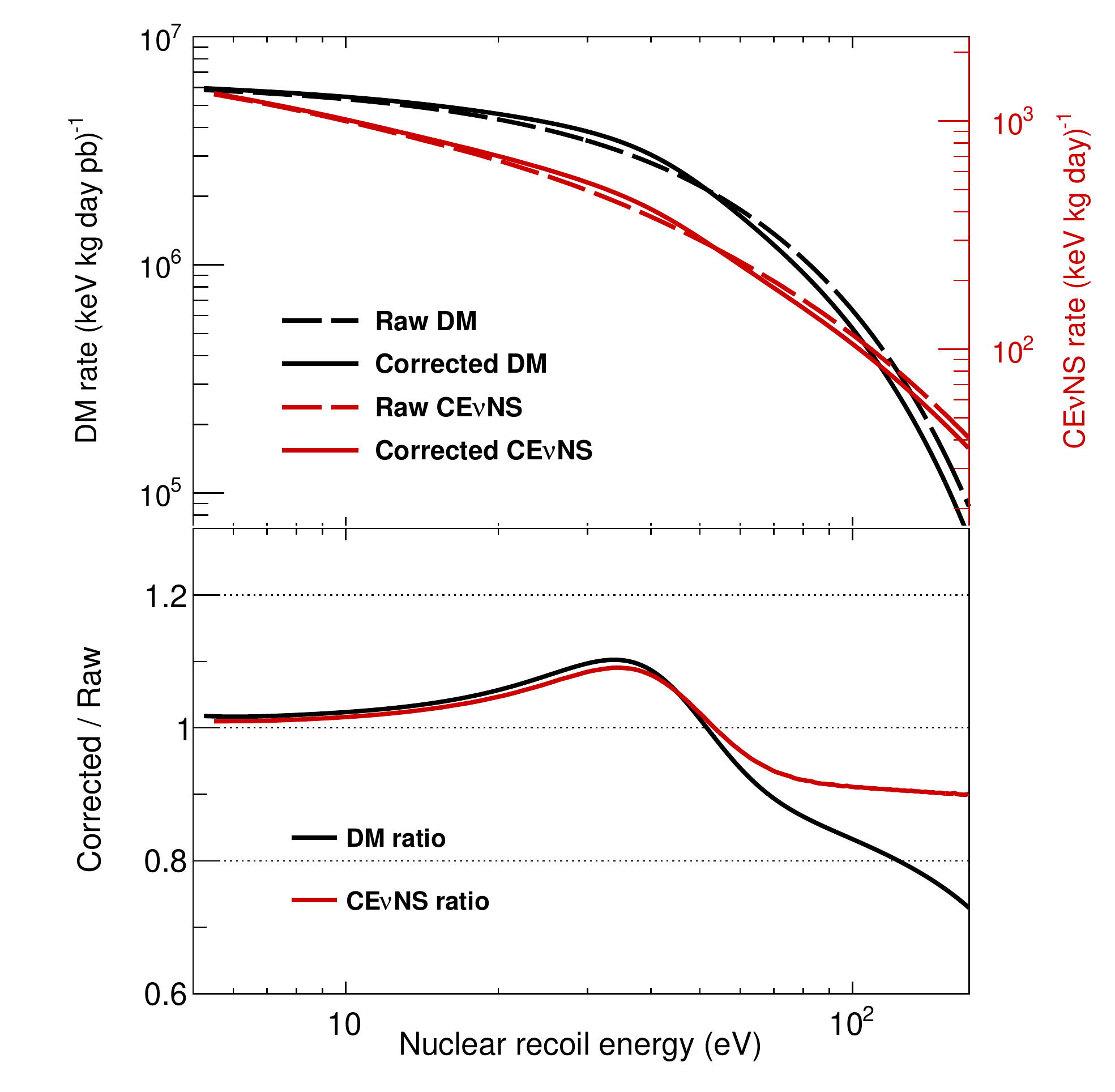}
    \caption{Expected nuclear recoil spectra from elastic scattering of 2 GeV DM particles (black curves) and reactor neutrinos (\CEvNS, red curves) with (plain lines) and without (dashed lines) taking into account the effect of lattice defects. The DM rate is normalized to the (unknown) DM-nucleon interaction cross-section.}
    \label{fig:cevns_dm}
\end{figure}

Then we apply our correction procedure to the spectra expected from the coherent elastic scattering of 2~GeV DM particles and reactor neutrinos (\CEvNS events). The raw and corrected recoil energy spectra are shown in figure \ref{fig:cevns_dm}. 
The calculation of the differential rate of DM-induced recoils was performed under the assumptions of a spin-independent interaction and standard halo model \cite{DONATO1998247}, while the \CEvNS spectrum is taken from \cite{NUCLEUS:2019igx}. Because of the coherent interaction with the target nucleus in both processes the detector response is dominated by far by the W atoms and we neglect the contribution of the other atoms. The ratio of corrected to not-corrected DM spectra show sizeable shape distortions, qualitatively comparable to the results published in \cite{PhysRevD.106.063012} for other materials. The ratio of the \CEvNS spectra is found to be similar to the 2GeV DM ratio in shape and amplitude. The relative distortions expected above a typical detection threshold of 20~eV cover a range of [$-$10, +10]\%. The impact on the shape of the detected spectrum could be degenerate with distortions expected from new physics scenarios, such as the contribution of a non-standard value of the neutrino magnetic moment \cite{Billard_2018,Papoulias:2019xaw}. In view of the planned high-precision measurements, the \CEvNS energy spectra will have to be corrected for the impact of crystal defects for an unbiased comparison with data.

We have reported the original calculation of the energy stored in lattice defects induced by $\mathcal{O}(10\--100)$\,eV recoils of tungsten atoms in \CaWO crystals. An interatomic linear ML potential has been designed from a large database of atomic configurations computed in the DFT framework . Using this quantitative potential, we calculated with MD, low energy recoils of tungsten atoms.
The energy range and temperature have been chosen for direct application to the prediction of recoil spectra in \CaWO cryogenic detectors. Beyond the first order $\approx$~10\% of energy lost in crystal defects, we predict a non-linear evolution of the average stored energy as a function of initial recoil energy due to the underlying distribution of displacement energy thresholds. The expected distortion of the continuous DM and \CEvNS spectra are given for reference. Taking the energy stored in crystal defects into account will be crucial for accurate interpretation of measured spectrum shapes. On a shorter term, accurate measurement of the mono-energetic tungsten recoils induced by thermal neutron capture could offer a unique probe of the lattice deformations we have calculated. 

\section{Acknowledgements}

The authors wish to thank Margarita Kaznacheeva for providing us with the calculated DM spectra and Anruo Zhong for providing helpful insights into the \MILADY package.

\bibliography{./references}{}

\end{document}